\documentclass[12pt,a4paper,twocolumn]{article}

\usepackage[utf8]{inputenc}
\usepackage[T1]{fontenc}
\usepackage{graphicx}
\usepackage{amsmath,amssymb}
\usepackage[numbers,sort&compress]{natbib}
\usepackage[hidelinks]{hyperref}
\usepackage{url}
\usepackage[margin=1.25cm]{geometry}
\usepackage{xcolor}
\DeclareRobustCommand{\investigator}{investigator}
\DeclareRobustCommand{\Investigator}{Investigator}
\title{Agentic MR method development and MR autoresearch: leveraging LLMs with MR skills for automatic physics-informed sequence and reconstruction development}

\author{
  \small
  Moritz Zaiss\textsuperscript{1,2}
  \thanks{Corresponding author: moritz.zaiss@uk-erlangen.de},
  Simon Weinm\"uller\textsuperscript{1},
  Amr Aly\textsuperscript{1},
  Lucas Kachelrie\ss\textsuperscript{1,3},
  \\
  \small
  Tobias Dornstetter\textsuperscript{1},
  Jonathan Endres\textsuperscript{1},
  Arnd Doerfler\textsuperscript{1},
  Andreas Maier\textsuperscript{2,3}
}

\date{%
  \vspace{-0.5em}\small
  {\setlength{\fboxsep}{6pt}\fbox{\parbox{0.92\linewidth}{\centering
    This article has been accepted for publication in \textit{Magnetic Resonance in Medicine} (2026).\\
    DOI: \href{https://doi.org/10.1002/mrm.70712}{https://doi.org/10.1002/mrm.70712}}}}%
}

\begin{document}

\onecolumn
\maketitle
\vspace{0.75em}
\setlength{\fboxsep}{10pt}
\noindent
\begin{minipage}[t]{0.288\linewidth}
\footnotesize
\raggedright
\setlength{\rightskip}{0.5em}
\textsuperscript{1}Institute of Neuroradiology, University Hospital Erlangen, Erlangen, Germany\\[0.4em]
\textsuperscript{2}Department of Artificial Intelligence in Biomedical Engineering, Friedrich-Alexander-Universit\"at Erlangen-N\"urnberg, Erlangen, Germany\\[0.4em]
\textsuperscript{3}Pattern Recognition Lab, Friedrich-Alexander-Universit\"at Erlangen-N\"urnberg, Erlangen, Germany\\[1.2em]
\textbf{Correspondence:}\\
Moritz Zaiss\textsuperscript{1,2} \raggedright
\href{mailto:moritz.zaiss@uk-erlangen.de}{moritz.zaiss@uk-erlangen.de}\\[1.2em]
\textbf{Funding Information:}\\
This work was supported by the German Federal Ministry of Research, Technology and Space (BMFTR), grant number 16SV9585.
\end{minipage}\hspace{0.4em}\vrule\hfill
\colorbox{gray!15}{\begin{minipage}[t]{0.652\linewidth}
\small
\textbf{Purpose:} Programming MRI pulse sequences is time-consuming and requires deep expertise in MRI physics, sequence design, and hardware constraints; even small modifications often require substantial debugging and validation. LLMs can assist when given structured prompts and error feedback, but many generated sequences still exhibit physical inconsistencies. We present Agent4MR, an agent-based framework that automatically generates and refines PyPulseq sequences using a structured, physics-aware validation report, and use it for autonomous research.\\
\textbf{Methods:} We evaluated Agent4MR on a spin-echo EPI task across three state-of-the-art LLMs and compared it to a context-only baseline and to a human developer with the same tools. We tested an autonomous research setup on a fluid-suppressed spin-echo EPI challenge for different model generations, in silico and in vivo. \\
\textbf{Results:} Across all models, Agent4MR consistently produced artifact-free, physically valid sequences in a single \investigator\ interaction, reducing the number of required interactions below the human baseline in a single comparison. Autonomous agents then accelerated and improved a sequence to match a given target contrast in an autonomous research (autoresearch) approach. In vivo, agents autonomously shortened a 6-shot reference scan by more than 50\%, keeping acceptable quality.\\
\textbf{Conclusion:} Embedding general-purpose LLMs in an agent loop with MR-specific context, code execution, and physics-based validation consistently produced valid spin-echo EPI sequences across several models. Competing agents also refined sequence and reconstruction against a defined target, in silico and in vivo. Agentic sequence development is therefore a promising path to reduce effort and expertise required to implement and iterate on MR imaging methods.\\[0.6em]
\textbf{Word count abstract/body:} $\approx$250 / $\approx$5022
\end{minipage}}
\vspace{1.5em}
\newpage
\twocolumn

\section{Introduction}

Programming MRI pulse sequences requires extensive research effort as well as deep expertise in MRI physics, hardware constraints, and sequence design, which makes it time-consuming. Moreover, even small modifications to existing sequences can involve non-trivial debugging and validation, creating a barrier to rapid prototyping and methodological innovation.

A massive leap to democratization of MR sequence programming came with the open-source MR sequence definition standard Pulseq and its Python implementation PyPulseq. These tools, for the first time, allowed MRI pulse sequences to be defined in a complete form outside of vendor-specific environments\cite{layton2017pulseq,ravi2019pypulseq}. This open, hardware-independent representation has enabled reproducible sequence development, sharing, and education. The combination of Pulseq with MR simulations provides a fast feedback loop for learning MR sequence programming\cite{zaiss2022mripulseq}. Importantly, this progress is fundamentally fueled by open MR sequence standards and frameworks such as Pulseq\cite{layton2017pulseq}, PyPulseq\cite{ravi2019pypulseq}, TOPPE\cite{nielsen2018toppe}, and gammaSTAR\cite{cordes2020gammastar}; without these approaches and publicly available code, little knowledge and few falsifiable tools would be available for training and testing language-model-based agents.

In parallel, recent advances in large language model (LLM) technology have demonstrated that these models are capable of generating complex code across a wide range of domains. Given these developments, it became conceivable that LLMs could potentially be applied to the complex task of writing valid MRI sequence code directly, leveraging both their coding abilities and access to comprehensive, open-source MR frameworks.

However, despite the vast capabilities of LLMs, there are still significant challenges in applying them to MRI sequence programming. One of the main hurdles is the need for a model to understand and reason about the complex physical constraints of MRI, such as the effects of RF pulses and their exact timing on tissue magnetization and contrast, as well as the impact of gradient waveforms and thus k-space sampling trajectories on image quality. These challenges require not only the ability to generate code but also the ability to reason about the physical implications of the code, which is a significant challenge for current LLMs.
Recent studies have demonstrated that LLMs can assist MRI sequence programming when guided by carefully structured prompts and iterative error feedback\cite{zaiss2024gpt4mr}. Prior work showed that concise, example-driven prompts can enable LLMs to generate executable PyPulseq-based MRI sequences, positioning LLMs as intelligent assistants rather than fully autonomous designers\cite{zaiss2025llm4mr}. That context-only baseline, a short MRI- and PyPulseq-specific prompt with example code and sequence-design rules, but no execution tools, is what we call LLM4MR below; the full context is given in\cite{zaiss2025llm4mr}. However, many generated sequences still exhibit physical inconsistencies, particularly in complex acquisitions. Building on these findings, we investigate whether an agent-based architecture with structured validation feedback can improve robustness for MRI pulse sequence generation using the PyPulseq library.

\subsection*{Agent4MR \& MR autoresearch}

\begin{figure}[htbp]
  \centering
  \includegraphics[width=\columnwidth]{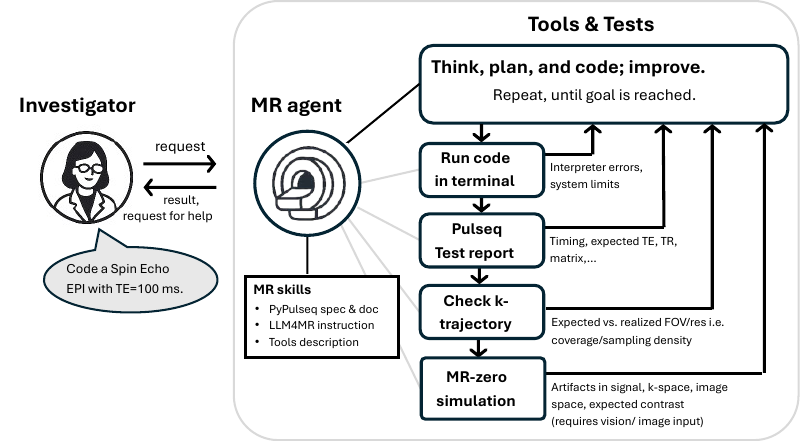}
  \caption{Overview of the \textbf{Agent4MR} framework, a Large Language Model with specific MR and PyPulseq knowledge context, tools and tests. The agent generates MRI pulse sequence code using PyPulseq, executes the sequence, and receives reports covering echo time (TE), repetition time (TR), k-space trajectory, gradient and RF raster timing, and other metrics. The agent iteratively refines the sequence until all constraints are satisfied, producing a scanner-ready, physically valid sequence.}
  \label{fig:overview}
\end{figure}

Herein, we present \textbf{\textit{Agent4MR}}, an agent-based framework that iteratively generates and refines pulse sequences using validation reports, which are provided to the agent after each execution cycle. An agent, in this paper, is a language model that writes code, runs it, and reads the validation report. This approach was first presented at ESMRMB~2025\cite{zaiss2025agentic}. An overview of the Agent4MR workflow is shown in Figure~\ref{fig:overview}.

The Agent4MR can then work autonomously on sequence and reconstruction approaches, which we tested in an \textbf{\textit{MR autoresearch}} setup where agents were asked to outperform each other, leading to better and better methods with regard to the given sequence and reconstruction task. Autoresearch, introduced by Karpathy\cite{karpathy2026autoresearch}, means that several agents attempt one task on their own and are ranked by a single score.

\section{Methods}
\subsection*{Agent4MR - LLM with MR-specific skills, validation tools and tests}

We extend the prompt-based workflows\cite{zaiss2025llm4mr} with an agent-based refinement loop (Agent4MR). The LLM operates in an agent mode, generating code and refining it based on structured feedback. We implement an automated validation report that analyzes a generated PyPulseq sequence and extracts key properties: echo time (TE), repetition time (TR), k-space trajectory characteristics, gradient and RF raster timing, and other consistency metrics. The agent continues refinement until the validation report indicates that constraints are satisfied or a predefined iteration limit is reached.

We evaluated all methods using a single metric: the number of \investigator\ interactions required until the \investigator\ was satisfied with the generated MRI pulse sequence. The investigator is the person who set the task, whether the other party is a model or the human developer. An interaction is a correction after an unsatisfactory output; fewer interactions mean more autonomous sequence generation.
The investigator accepted a result when the image had no gross artifacts and the spin echo coincided with the acquisition of the k-space center. A small residual misalignment between odd and even echoes, visible as a weak Nyquist ghost, was still accepted. The CPMG phase condition was not required, because with a single refocusing pulse it only changes the global image phase.
The corrections were short hints, for example ``the spin echo is not centered in the ADC'' or ``the prewinder has the wrong polarity''.
The ideal case requires only the initial task prompt.
All experiments used a similar target task:

\begin{quote}
``Please, read the instructions.\\ Code a spin echo EPI sequence (64×64, FOV\,=\,(0.2, 0.2, 0.008)\,m, TE\,=\,100\,ms).'' 
\end{quote}

For each of three base LLMs (Gemini~2.5 Pro, GPT-5, Claude~Opus~4.1), we compared three setups: (1)~LLM - direct instruction without any specialized context or tools; (2)~LLM4MR - instruction using a concise, MRI-specific and PyPulseq-specific context, but no tools; (3)~Agent4MR - our agent-based system with automated execution and validation feedback. 
Each setup was applied five times; we report the mean and standard deviation of the number of required \investigator\ interactions. 

The agent system is implemented in Python; orchestration, tool calling, and validation feedback loops are handled programmatically. Sequence generation and execution use the PyPulseq library (v1.4.2). Cursor AI is used as the environment for code generation, execution, and feedback ingestion. We evaluated the system using a standardized task: implementing a spin-echo EPI sequence with TE=100\,ms, with the goal of generating \texttt{.seq} files across different base LLMs that yield artifact-free images. The evaluation prompt was: 

\begin{quote}
``Please, read the Agent4MR instructions. \\Code a spin echo EPI sequence (64×64, FOV\,=\,(0.2, 0.2, 0.008)\,m, TE\,=\,100\,ms). Finish only when you are 100\% sure it is properly implemented. Use the terminal to test the codes. Use all tools to validate the sequence.'' 
\end{quote}

As a comparable human baseline we chose a student assistant developer who was recently trained in PyPulseq programming and already solved other sequence tasks in our research group, but was at the same time naive to the spin echo EPI concept. The human was prompted with the identical task allowing the same tools available as for the Agent4MR.

\subsection*{MR autoresearch - An Agent4MR research community}

Following Karpathy's autoresearch approach\cite{karpathy2026autoresearch}, we formulate a research goal and let multiple agents try to solve it autonomously. In the original autoresearch, an AI agent repeatedly edits a machine-learning training script, runs a short experiment, measures the result, and keeps only changes that improve a fixed score, so that many experiments run without human intervention. We transfer this idea to MR sequence development: the editable code is the PyPulseq sequence and its reconstruction, the experiment is an MR simulation or scan, and the score is an image-quality metric. A leaderboard is the shared list that ranks each attempt by this score. In this way, the agents form a small research community in which successful ideas are reused and combined and failed ones are documented, while the investigator only defines the task, the constraints, and the score.
For ranking on a leaderboard we need a focused and simple research topic, with a clear quality metric. Following our initial Spin Echo EPI experiments, which were chosen for their simplicity in code but reasonable complexity in MR physics, we propose improving a baseline FLAIR spin echo EPI under the constraint of less than 10\,s of scan time.

This forms a more complicated task where contrast, distortions and scan time have to be maintained simultaneously.
To rank the attempts we need a clear quality metric, for which we choose the mean absolute error (MAE) with regard to an ideal signal equation image

\begin{equation}
\label{eq:1}
S_{\mathrm{tgt}}(\vec{r})
=
\left|
\rho(\vec{r})\,e^{-\mathrm{TE}/\mathrm{T}_2(\vec{r})}
\left[1-2e^{-\mathrm{TI}/\mathrm{T}_1(\vec{r})}\right]
\right|,
\end{equation}

with TI being the inversion time after a 180° pulse, and $\mathrm{T}_1$, $\mathrm{T}_2$, as well as the proton density $\rho$ being the brain phantom $P$'s property maps.
This forms an ideal FLAIR SE EPI without any $B_0$ or $B_1^+$ related artifacts, and no limitation from truncated frequency and phase Fourier encoding.
The signal equation does not include $\mathrm{T}_2^*$ blurring or distortion along the echo train. This is intentional: the target is the ideal image, and mitigating these effects by sequence and reconstruction design is part of the challenge.
In contrast, the simulated image is generated by

\begin{equation}
\label{eq:2}
S_{\mathrm{tsk}}(\vec{r})
=
\left|
\mathbf{reco(sim}\!\left[\mathbf{seq}(\mathrm{TE},\mathrm{TI}),P\right])(\vec{r})
\right|,
\end{equation}

with $\mathbf{reco}$ being the reconstruction operator. $\mathbf{sim}$ denotes the MR-simulation operator governed by the sequence $\mathbf{seq}$, acting on all brain phantom properties $P$, including typical inhomogeneous $B_0$ and $B_1^+$, as well as $\mathrm{T}_2'$ and isotropic diffusion effects. The agents can adjust the code of both the PyPulseq sequence definition, as well as that of the reconstruction module code. Overfitting to the target, too long scan time (> 10\,s) or too long reconstruction time (> 2\,min) leads to disqualification.
All other attempts are rated successful and can be written into a leaderboard.md sorted by the MAE with regard to the ideal target. 

\begin{figure}[h]
  \centering
  \includegraphics[width=\columnwidth]{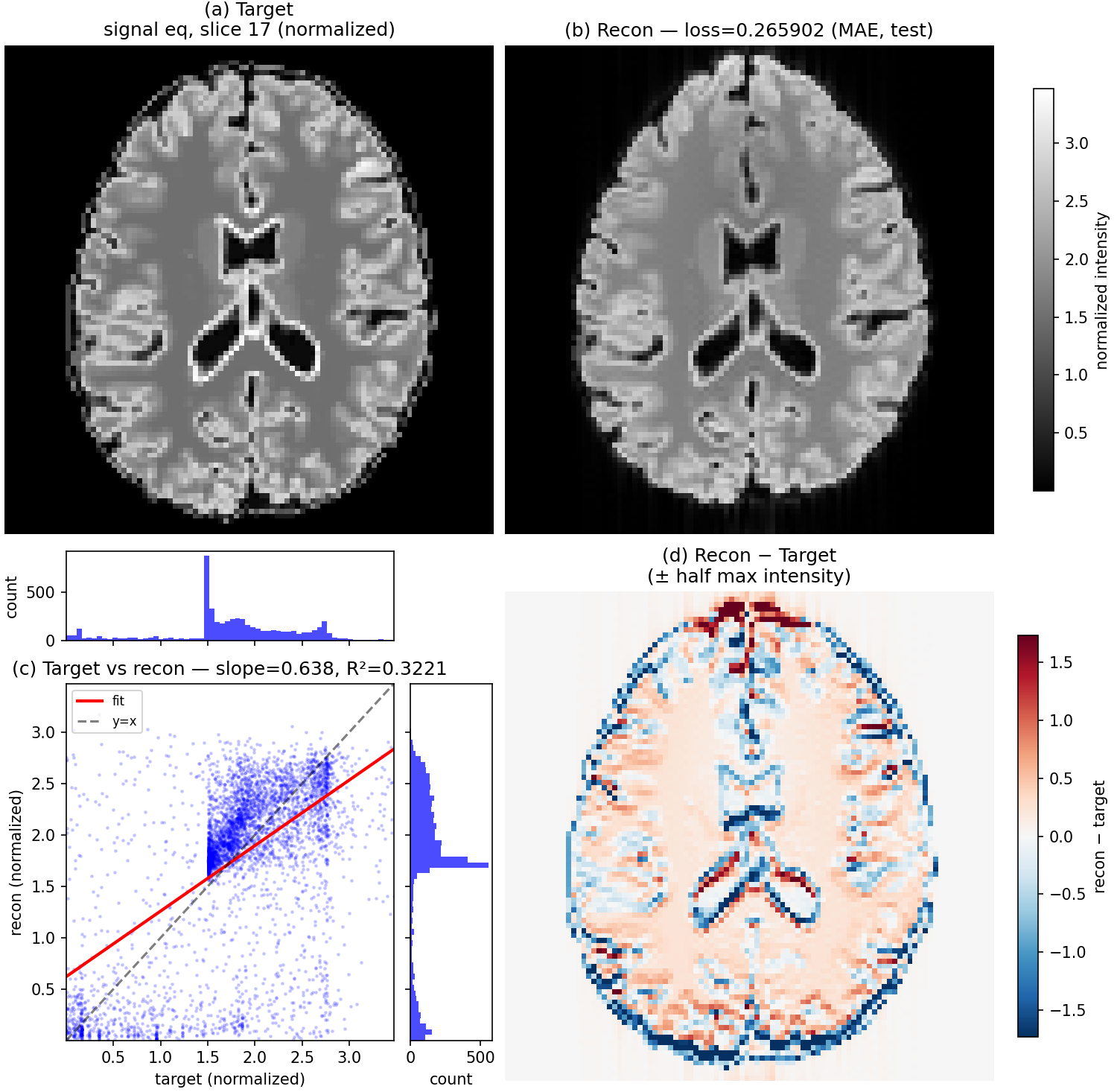}
  \caption{Comparison of (a) the target defined by the signal equation (eq. \eqref{eq:1}) and (b) the simulation result (eq. \eqref{eq:2}) of the initial sequence. (c) shows the regression plot with histograms and (d) the difference image. The actual challenge is to mitigate EPI image distortions, while achieving the right contrast, as well as staying below 10\,s of scan time and scanner hardware limits.}
  \label{fig:MRauto_000}
\end{figure}

We provide the agents with initial code for a 2D $96\times96$ single shot inversion recovery spin echo EPI with TE\,=\,100\,ms and TI\,=\,2.8\,s, which has an MAE of 0.27. It is fully sampled and not optimized for bandwidth or exact timing. The comparison of target and the image created by the initial code can be seen in Figure \ref{fig:MRauto_000}, including distortions and contrast deviations in the MR sequence output.

We do not simulate slice selection effects nor gradient imperfections or random phases between shots.
As simulation intensity can vary with resolution, we calculate the MAE after scaling the images to a unit mean, isolating the image contrast. The signal intensity or SNR is thus not enforced. We also have no SAR restrictions, phantoms are static, and the setup relies on a single-channel coil, thus no parallel imaging capability was available for acceleration. Furthermore, we used an MR system with maximal gradient amplitude of 40\,mT/m and a slew rate of 150\,mT/m/s.

The instructions for the leaderboard challenge were given as a Markdown file to the agents; the direct prompt was then simply:

\begin{quote}
``Please, read the instructions.\\Approach to win the challenge!'' 
\end{quote}

All agents write and run in the same challenge folder, read \texttt{leaderboard.md}, and can open earlier numbered experiment folders and reuse their code. In addition to the above described Agent4MR tools, we added a skill describing how to use MR-zero\cite{Loktyushin2021} for end-to-end optimization. 

\subsection*{In vivo MR autoresearch}

For in vivo MR autoresearch, we extended the Agent4MR loop so that agents can send Pulseq sequences to a real scanner, load multi-coil raw data, and reconstruct images with the same reconstruction codes used in simulation (extended for coil combination). Real-scan experiments were performed on a whole-body 3T scanner (CIMA.X, Siemens Healthineers) using a 20-channel head coil. One healthy subject was scanned after written informed consent, within a study approved by the local ethics committee. All candidate sequences passed the Pulseq sequence tests before playout, and the scanner's SAR and peripheral nerve stimulation (PNS) monitoring remained active, ensuring subject safety as for any Pulseq sequence.

Because an analytic signal-equation target is not available on the scanner, we fixed a \textbf{measured 6-shot IR-SE-EPI} acquisition as the target image (matrix $128\times128$, $\mathrm{TI}=2.2$\,s, recovery time $T_{\mathrm{rec}}=10$\,s, one dummy shot, $\mathrm{TE}=100$\,ms, slice thickness 10\,mm; reference duration $t_{\mathrm{ref}}\approx86$\,s). Candidates may use any sequence type and any reconstruction, but must return a magnitude image of shape $(128,128)$. Each evaluation reacquires the 6-shot reference and the candidate back-to-back in the same session, then scores the candidate against that fresh measured reference. This reduces the mismatch between reference and candidate caused by subject motion or system drift between scans. A parallel simulation track on a 20-coil numerical phantom (same fixed 6-shot target) is used for development test in simulation. The final ranking is decided on real-scan metrics.

Because a near-perfect match is always possible by simply reproducing the long reference, the objective is the \emph{fastest} sequence that still approaches the target. We therefore rank by a combined loss that forms a compromise between accuracy and duration,
\begin{equation}
\label{eq:realscan_qs}
\begin{aligned}
\mathrm{loss}&=1-q\cdot s,\\
q&=\frac{\mathrm{MAE}_{0}-\mathrm{MAE}}{\mathrm{MAE}_{0}},\\
s&=\frac{t_{\mathrm{ref}}-t}{t_{\mathrm{ref}}}
\end{aligned}
\end{equation}

where $\mathrm{MAE}_{0}$ is the baseline single-shot real-scan error ($\approx0.195$), $t$ is the candidate scan duration, and $t_{\mathrm{ref}}$ is the duration of the measured 6-shot IR-SE-EPI reference ($\approx86$\,s). Lower loss is better ($0$ would correspond to a perfect, instantaneous match). Candidates with a scan duration longer than the reference ($t>t_{\mathrm{ref}}$) are rejected. A perfect match gives $q=1$ and thus $\mathrm{loss}=1-s$, which still depends on the duration and is zero only for $t=0$. A candidate without improvement over the baseline ($q=0$) has $\mathrm{loss}=1$, independent of its duration.

\section{Results}
As defined above, we report the number of \investigator\ interactions required until the \investigator\ was satisfied with the generated MRI pulse sequence; lower values indicate more efficient and autonomous sequence generation (task prompts, see Methods and Supporting Table~\ref{stab:prompts}).

\begin{figure*}[tbp]
  \centering
  \includegraphics[width=0.95\textwidth]{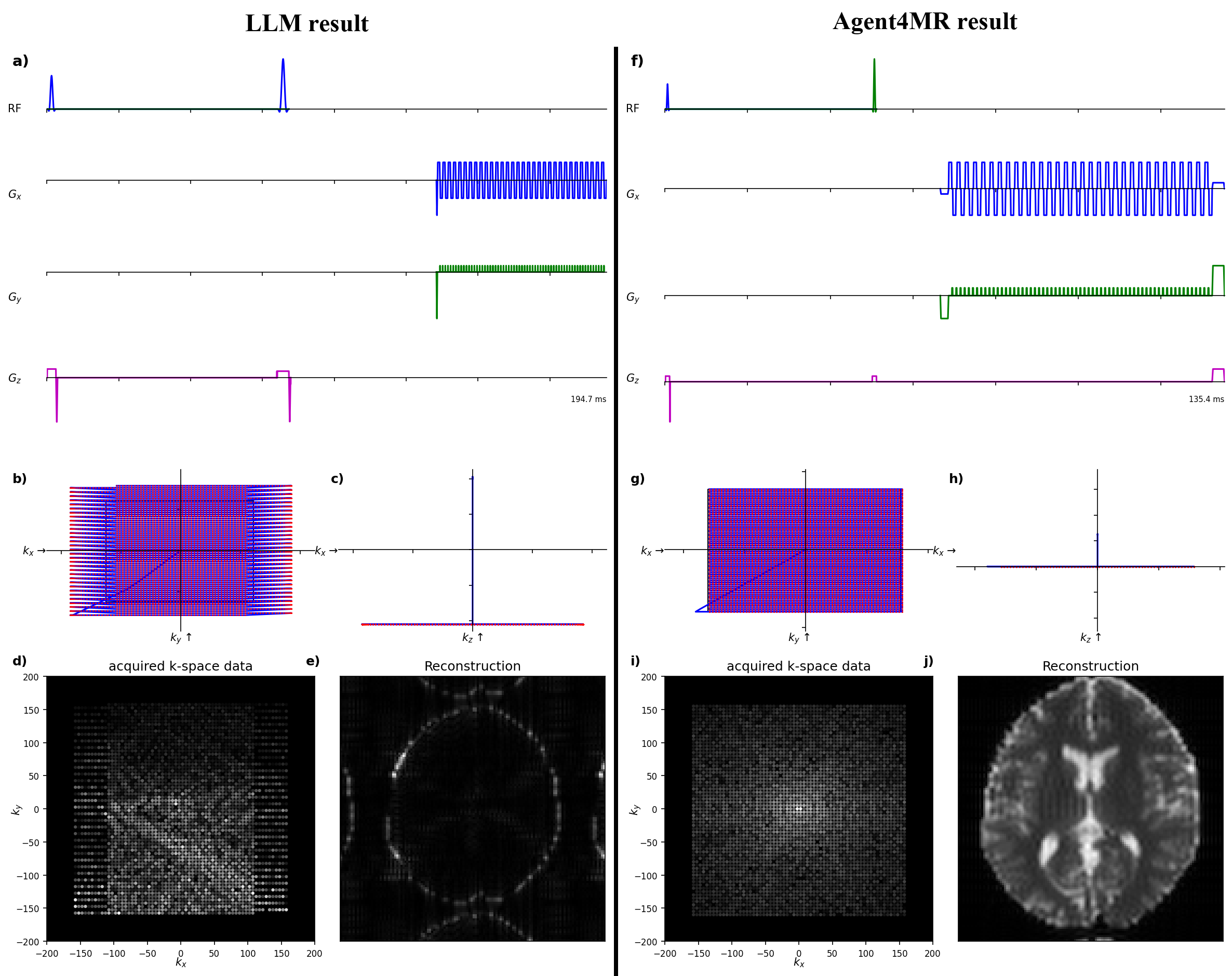}
  \caption{Example output from a bare LLM (a--e) and from Agent4MR (f--j) for the spin-echo EPI task stated in Methods. The bare LLM makes several careless mistakes: a phase rewinder after the refocusing pulse, leading to remaining z-dephasing visible in c) and to signal cancellation in e). A wrong x-prewinder, leading to wrong k-space coverage in b), leading to FOV error in e), and a wrong echo time. The Agent4MR correctly implemented all these features (f), leading to correct k-space coverage (g--i) and image (j).}
  \label{fig:fig2_bare_llm_seq}
\end{figure*}

Figure~\ref{fig:fig2_bare_llm_seq} shows an example of a generated sequence of a bare LLM and the resulting artifact-affected simulated image for this task. The typical finding was that most necessary building blocks of the sequence were present, but the LLM had difficulty in placing them in the correct order and with the correct parameters without further context.

\begin{figure*}[htbp]
  \centering
  \includegraphics[width=0.95\textwidth]{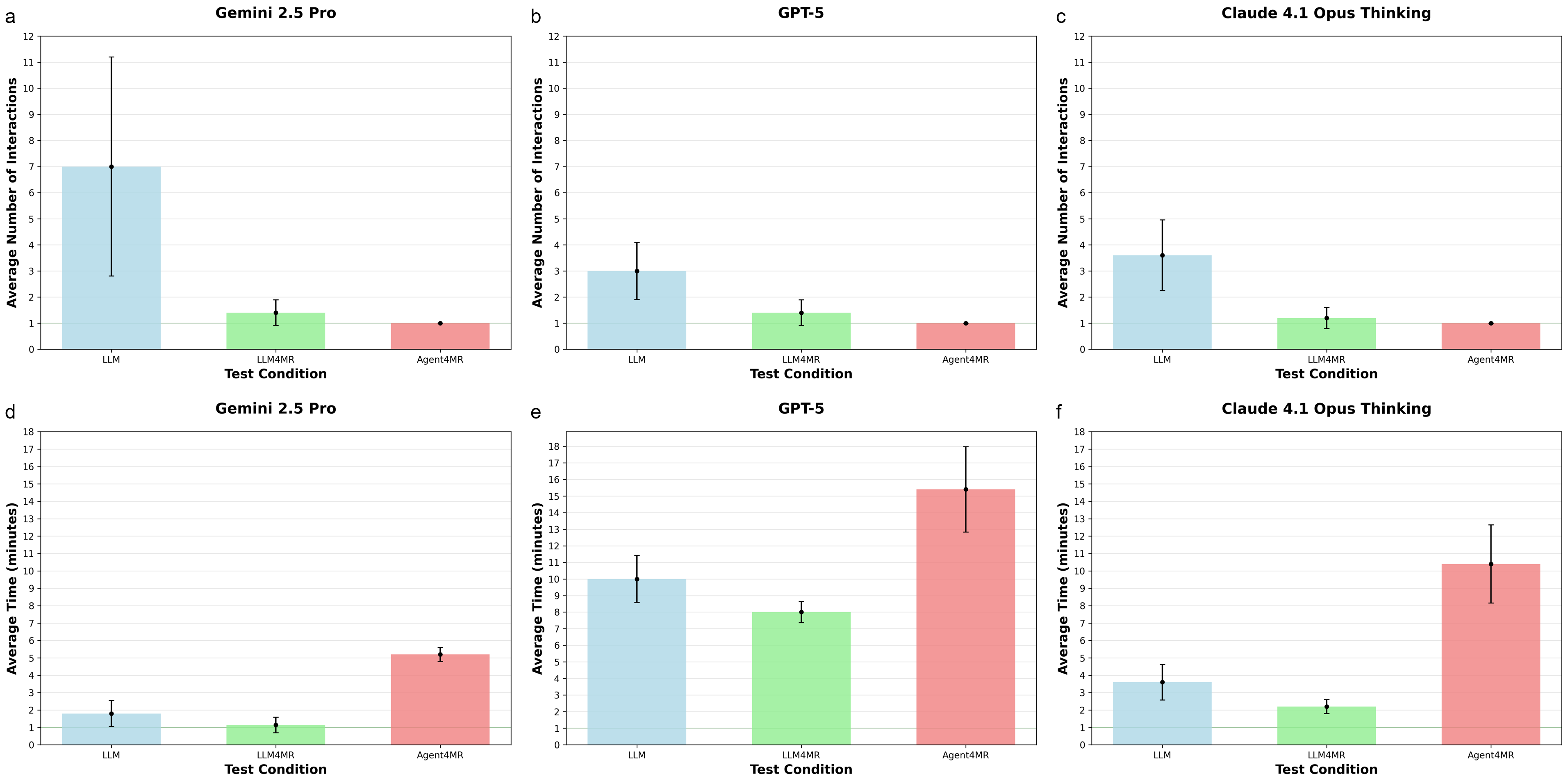}
  \caption{Average number of \investigator\ interactions per condition for each base LLM (a--c). Average time required by each agent to finish the request (d--f). Time for generation of tests and simulation of sequences is included, single simulation time is below 10\,s.}
  \label{fig:results_combined}
\end{figure*}

Figure~\ref{fig:results_combined}(a--c) summarizes the results per model for the three setups (LLM, LLM4MR, Agent4MR).

\subsection*{Effect of context and tool-equipped agent refinements on \investigator\ interactions}

Across all evaluated models, the use of the concise MRI-specific context (LLM4MR) substantially reduced the number of required \investigator\ interactions compared with the unprompted baseline (LLM), see blue to green bar improvement in Figure~\ref{fig:results_combined}(a--c). This trend was consistent across Gemini~2.5 Pro, GPT-5, and Claude~Opus~4.1. Without prompting, models required multiple corrective interactions to resolve syntax errors, incorrect PyPulseq usage, or missing sequence components. Introducing the MR- and PyPulseq-context improved task grounding and reduced both the mean interaction count and its variability, indicating more stable behavior across repeated trials.
However, even with the LLM4MR context, many LLM-generated sequences exhibited errors in key parameters such as TE, TR, k-space trajectory, or gradient timing: LLM-only sequences frequently showed physically inconsistent trajectories or incorrect timing, while LLM4MR sequences improved but still contained residual errors. In contrast, the Agent4MR configuration generated sequences that were consistently correct, satisfying the target acquisition parameters, k-space trajectories, and timing constraints without \investigator\ intervention, see green to red bar improvement in Figure~\ref{fig:results_combined}(a--c). The agent's iterative refinement loop, driven by the structured validation report, allowed it to autonomously detect and correct errors in TE, TR, and gradient events, producing fully valid sequences across all evaluated LLMs. 

This comparison confirms that the agent framework is robust across LLMs, which indicates that for a well-defined coding task such as this one, the context and validation might be more important than the exact model. For open-ended optimization, later model generations performed clearly better, from Gemini~2.5 Flash over Gemini~3~Flash to Gemini~3.1 Pro (Figure~\ref{sfig:2}, Figures~\ref{fig:MR_autoresearch_fig1} and~\ref{fig:MR_autoresearch_g31}). All models were used in thinking mode, which means they have intrinsic ability to reason about the task and the constraints. This is necessary for tool use but also increases the duration and number of required tokens for each model, see Figure~\ref{fig:results_combined}(d--f) and Table 1. Gemini was generally the fastest model, solving the task in under 6 minutes. Thus, the underlying model capabilities still influence the results in solution time and also with regard to sequence fidelity below the currently used threshold.

\begin{table}[htbp]
  \centering
  \caption{\Investigator\ interactions N, time, token usage, and estimated cost until a satisfactory sequence. Human baseline: recently trained EPI-naive student, 2.5\,h, 5 interactions.}
  \label{tab:cost}
  \begin{tabular}{lccrr}
    \hline
    Configuration & N & Time & Tokens & Cost [\texteuro]  \\
    \footnotesize{Agent4MR} & & \footnotesize{[min]}  & & \footnotesize{[$\approx$]}\\
    \hline
        \small{Gemini~2.5 Pro} & 1 & 4.8 & 87k & 0.12 \\
    \small{GPT-5}  & 1 & 15 & 520k & 0.48 \\
    \small{Claude~Opus~4.1} & 1 & 10.2 & 1162k & 10.00 \\
    \small{Human} & 5 & 150 & --- & 37.50  \\
    \hline
  \end{tabular}
\end{table}

Finally, we wanted to compare the agentic framework to a human baseline, using the same task. The human completed the task in approximately 2.5\,h with five interactions with the \investigator. Table~\ref{tab:cost} summarizes the results. The results show that the agentic framework is much faster and requires fewer interactions than the human baseline, but also that the human baseline is not as bad as the LLM-only baseline.
In terms of cost, the agentic framework was for this task significantly cheaper than the human baseline with a wage of approx 15\,\texteuro{} per hour assumed for a student assistant. Only the token expenses during the run of the experiment are taken into account. 

We also tested these agents in larger projects where sequence defining code has more dependencies, submodules and more general complexity. Also here tasks could be resolved with only one interaction, e.g. adding to an existing full feature FLASH sequence a multi-echo functionality with both bipolar and monopolar gradient options, as well as echo labels in PyPulseq (data not shown).

\subsection*{Autonomous MR research agents}

\begin{figure*}[!ht]
  \centering 
  \includegraphics[width=1.0\textwidth]{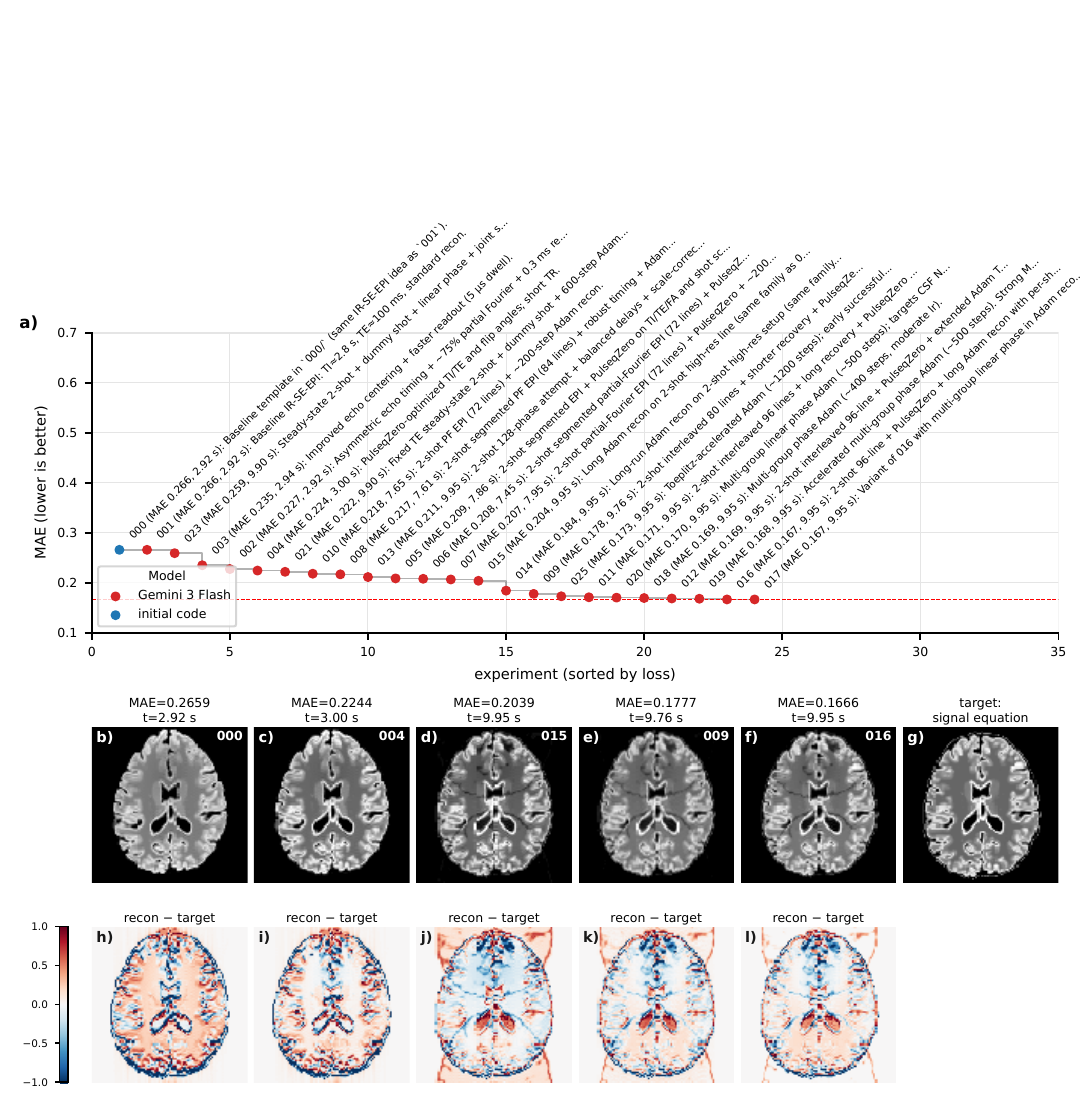}
  \caption{IR-SE-EPI MAE leaderboard progress for Gemini~3~Flash. Autonomous agents iteratively improved the IR-SE-EPI pipeline (sequence parameters, reconstruction, post-processing). (a) Experiments are ranked by MAE loss (worst to best). The staircase shows improvement from baseline ($\sim$0.2659) to best ($\sim$0.1666) through two-shot interleaved and partial-Fourier EPI, differentiable MR-zero optimization of TI, TE and flip angle, and an iterative reconstruction with per-shot phase correction. (b--f) Reconstructions of selected experiments, (g) signal-equation target, (h--l) corresponding difference images (recon $-$ target).}
  \label{fig:MR_autoresearch_fig1}
\end{figure*}

\begin{figure*}[!ht]
  \centering
  \includegraphics[width=1.0\textwidth]{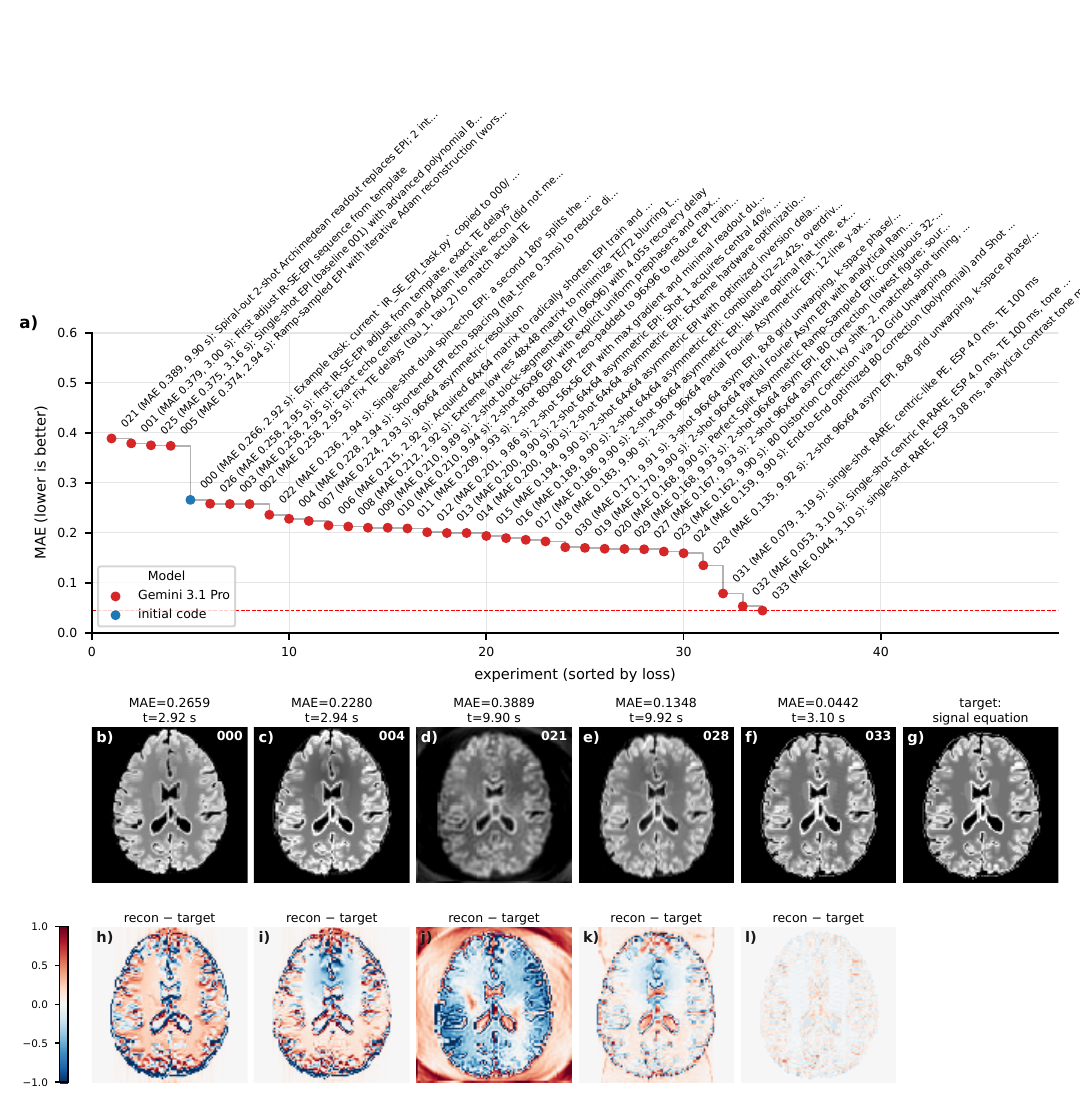}
  \caption{IR-SE-EPI MR autoresearch with the same setup as Figure~\ref{fig:MR_autoresearch_fig1} (Gemini~3~Flash), repeated with the later model generation Gemini~3.1 Pro over 34 experiments.
  Other attempts used ramp sampling, partial Fourier, lowering the resolution with upsampling, or spiral EPI readouts.
  The best spin-echo EPI reaches an MAE of $\sim$0.135, and a fast single-shot turbo spin echo reaches $\sim$0.044.
  Panels as in Figure~\ref{fig:MR_autoresearch_fig1}: (a) leaderboard, (b--f) reconstructions, (g) target, (h--l) difference images.}
  \label{fig:MR_autoresearch_g31}
\end{figure*}

Using the Agent4MR setup, we ran an autonomous agent competition with the goal to refine a spin-echo EPI sequence with a FLAIR preparation and a duration not exceeding 10\,s and its reconstruction to minimize the MAE between the simulated reconstruction and a signal-equation target. Each agent's sequence performance was noted in a leaderboard.

A summary plot of such a challenge, run with Gemini~3~Flash, is given in Figure~\ref{fig:MR_autoresearch_fig1}(a). The agents came up with multiple approaches to lower the bandwidth, and the EPI readout duration, and thus lower image distortions. At the same time the agents started grid search as well as differentiable optimizations using MR-zero\cite{Loktyushin2021,Endres2024}, to adjust the image contrast via TI and TE. Furthermore, adaptive reconstruction approaches were introduced to bring the mismatching signals of two shot sequences in coherence.
The FLAIR preparation together with the 10s time restriction is hard to solve perfectly, and also the Gemini~3~Flash agents did not find a near-perfect EPI solution. Still, the exploration and analysis by the agents consistently improved the images with regard to the given quality metric. 

Over 26 experiments, the agents improved from a baseline MAE of $\sim$0.266 (TI=2.8\,s, TE=100\,ms, single-shot 96×96 NUFFT) to a best MAE of $\sim$0.167. As each agent had access to the previous attempts the winning strategy combined multiple previous approaches:  2-shot EPI with per-shot signal scaling, MR-zero-optimized TI/TE/FA, and an iterative (Adam) reconstruction with per-shot phase correction instead of the NUFFT reconstruction of the baseline.
The same experiment was repeated with the earlier Gemini~2.5 Flash (Figure~\ref{sfig:2}) and the later Gemini~3.1 Pro (Figure~\ref{fig:MR_autoresearch_g31}), and later model generations created better sequences. Other attempts used ramp sampling, partial Fourier, lowering the resolution with upsampling, or spiral EPI readouts. Gemini~3.1 Pro now achieves a lower MAE than the human-designed sequence, both with a distortion-corrected two-shot SE-EPI and by switching to a fast single-shot turbo spin echo. 

Re-ranking these experiments by five further metrics, including the perceptual metric DISTS and the MAE after deformable registration, which removes geometric distortion (Figure~\ref{sfig:3}), shows that the metric matters for imperfect solutions: the intermediate EPI entries are ranked differently. Once an image comes close to the target, however, all metrics converge, and the single-shot TSE entries 032 and 033 are the two best entries in every metric.

Still, paths to the winner were not direct, but with failed attempts and several disqualified attempts that had longer than 10-s scan time. This challenge ran on one single PC within few hours, with a total token cost of about 13\,\texteuro. Individual experiments, including multiple simulation and reconstruction runs, could take up to 1h. Despite being encouraged to report also non-winning results, the agents were generally competitive and refined approaches to get a low MAE before publishing it to the leaderboard.

\begin{figure*}[htbp]
  \centering 
  \includegraphics[width=1.0\textwidth]{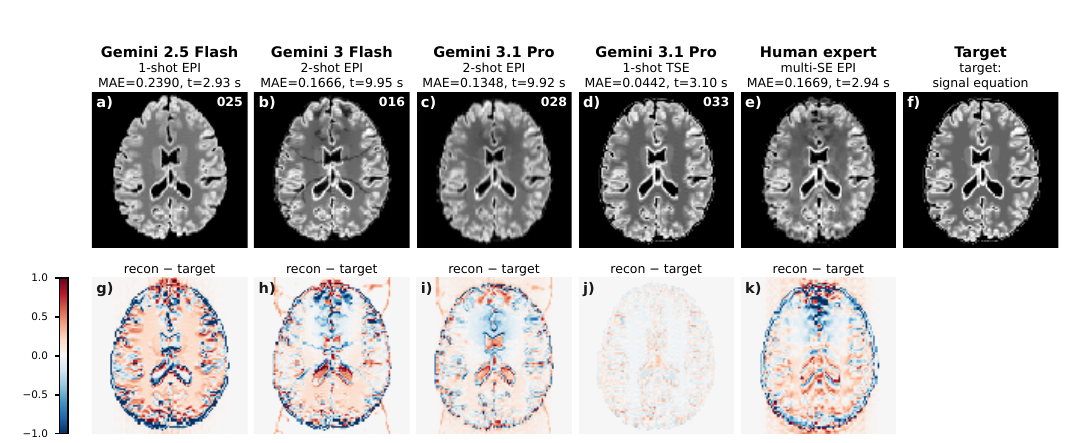}
  \caption{Overview of the in silico MR autoresearch winners. (a--d) Best entries of Gemini~2.5 Flash, Gemini~3 Flash and Gemini~3.1 Pro (best two-shot EPI and single-shot TSE), compared with (e) the human-invented and -implemented multi-SE EPI sequence with an MAE loss of 0.1669 and (f) the signal-equation target; (g--k) corresponding difference images (recon $-$ target). The human sequence had echo train length of 7, 5 k-space segments and two dummy refocusing pulses.}
  \label{fig:MRautoresearch_human}
\end{figure*}

Staying with the EPI approach this remains a tough challenge also for the human expert, who chose a multi-SE EPI after one FLAIR pulse. The results of this human-invented and -implemented sequence are shown  in Figure \ref{fig:MRautoresearch_human}. This sequence had a MAE loss of 0.1669, which interestingly was almost on par with the best result of Figure \ref{fig:MR_autoresearch_fig1} (MAE\,=\,0.1666), but was outperformed by Gemini~3.1 Pro both with the SE-EPI (MAE\,=\,0.135, $t=9.92$\,s) and the TSE (MAE\,=\,0.044, $t=3.10$\,s) approaches (Figure~\ref{fig:MR_autoresearch_g31}). The authors did not consider that a TSE could win the challenge because of the rather long echo train length of 96, but changing the sequence type was not forbidden in the challenge.  The most elegant and potentially better realization of such a multi-"SE-EPI" might be the spiral TSE of Hennig et al.\cite{henig_spiraltse}. Gemini~3.1 Pro indeed left the EPI concept with a single-shot TSE; a HASTE with variable flip angles or a T1/T2-prepared FLASH would be further candidates.

We also ran an early experiment without time restriction, which could be solved by a six-shot approach with near perfect result, see also Figure~\ref{sfig:1}. However, the six shot sequence won by using much more scan time, so we moved the challenge to the more interesting time restricted version.

\subsection*{Token Usage and Cost Analysis}

The experimental evaluation involved a significant volume of token processing across three model generations. As shown in Table \ref{tab:token-usage}, the Gemini 3 Flash model exhibited the highest throughput, for 25 evaluated experiments it processed over 136 million tokens. For comparison, 136 million tokens corresponds approximately to 102 million words, which is equivalent to about \textbf{17,000} research articles of 6,000 words each.  While Gemini 3.1 Pro processed fewer tokens than Gemini 3 Flash, it accounted for the largest portion of the experimental expenditure, totaling approximately 59.72\,\texteuro{} for 25 experiments. Overall, the experiments across all models required a total of 233.8 million tokens, with an estimated API cost of 75.00\,\texteuro.

Interestingly, the final codes the agents created for the challenge were approximately 3,000–5,000 tokens. Thus around 125,000 tokens for one challenge, still the context, the thinking processes and the tests required 1,000 times more tokens, compared to the final output. 

\begin{table}[ht]
\centering
\caption{Normalized Token Usage and API Costs in EUR (n=25 experiments per model)}
\label{tab:token-usage}
\begin{tabular}{crrr}
\hline
\textbf{Gemini} & \textbf{Total}  & \textbf{Tokens } & \textbf{Cost}\\
\textbf{Model} & \textbf{Tokens}  & \textbf{per Exp.} &   \\
               & \textbf{[M]} & \textbf{[M]} & \textbf{[\texteuro]}\\ \hline
\small{2.5 Flash} & 25.5 & 1.02 & 2.43 \\
\small{3 Flash} & 136.3 & 5.45 & 12.85 \\
\small{3.1 Pro} & 72.0 & 2.88 & 59.72 \\ \hline
\textbf{Total} & \textbf{233.8} & \textbf{9.35} & \textbf{75.00} \\ \hline
\end{tabular}
\end{table}

\subsection*{In vivo MR autoresearch}
\begin{figure*}[htbp]
  \centering
  \includegraphics[width=1.0\textwidth]{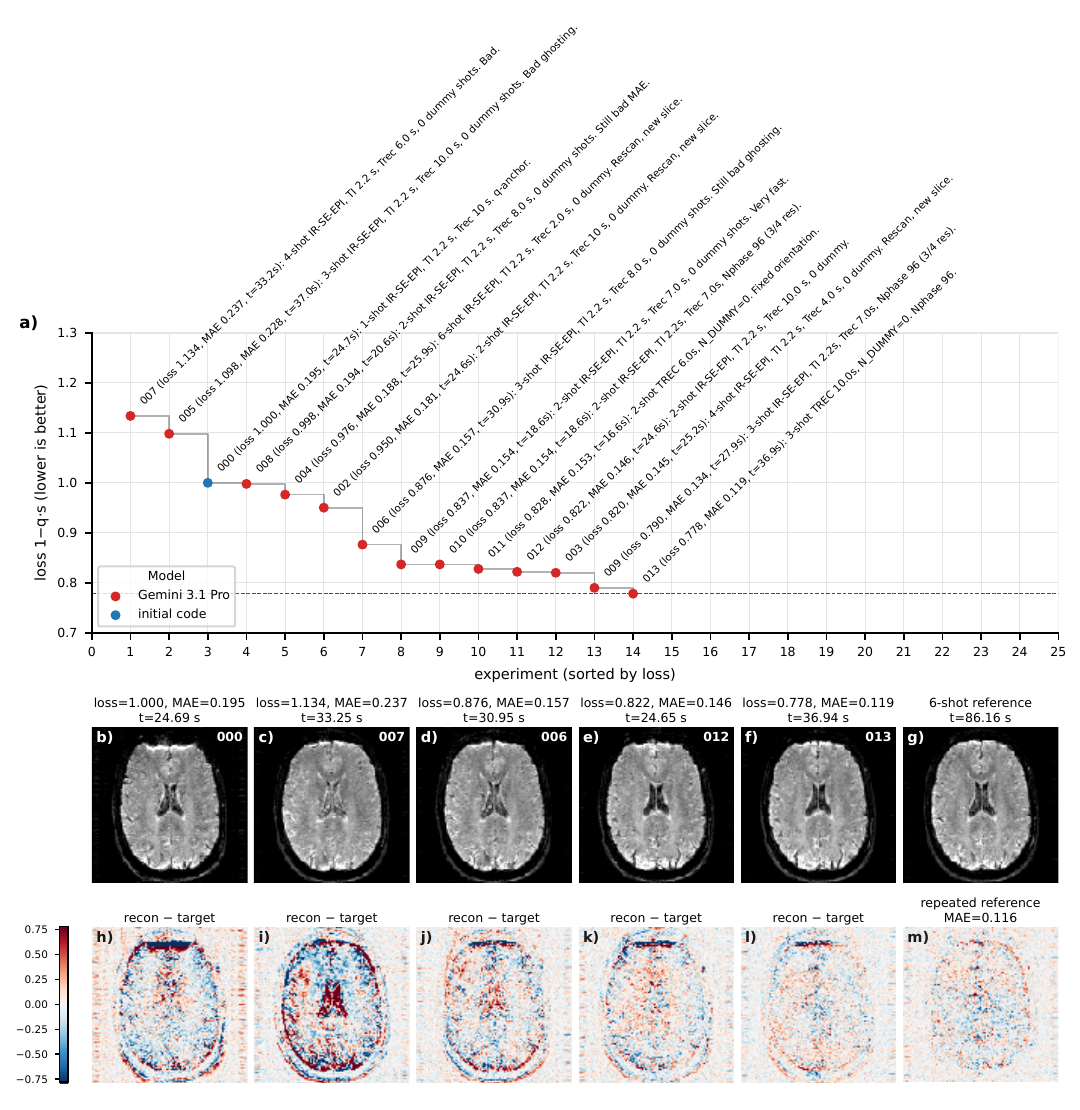}
  \caption{In vivo MR autoresearch on a 3T scanner. Ranking uses $\mathrm{loss}=1-q\cdot s$ (eq.~\eqref{eq:realscan_qs}) on in vivo MAE and duration (lower is better).
  (a)~Leaderboard progress: experiments sorted by decreasing loss.
  (b)~Baseline single-shot IR-SE-EPI: MAE$\approx0.195$, $t\approx25$\,s, $\mathrm{loss}=1.0$.
  (c)~A poorer 4-shot attempt: MAE$\approx0.237$, $t\approx33$\,s, $\mathrm{loss}\approx1.13$.
  (d, e)~Intermediate 3-shot and 2-shot IR-SE-EPI ($\mathrm{loss}\approx0.88$ and $0.82$).
  (f)~Best entry for this compromise, 3-shot IR-SE-EPI: MAE$\approx0.119$, $t\approx37$\,s, $\mathrm{loss}\approx0.78$.
  (g)~6-shot reference, $t_{\mathrm{ref}}\approx86$\,s.
  (h--l)~Difference images (recon $-$ target) of (b--f); (m)~difference of two repeated reference scans (MAE$\approx$0.116).}
  \label{fig:realscan_autoresearch}
\end{figure*}

In vivo, the autoresearch approach led to 13 reported attempts (Figure~\ref{fig:realscan_autoresearch}) generated and executed within a single subject session of about 1.5\,h. The best $1-q\cdot s$ entry was a 3-shot IR-SE-EPI ($\mathrm{MAE}\approx0.119$, $t\approx37$\,s; $\mathrm{loss}\approx0.78$), improving on the single-shot baseline (MAE$\approx0.195$) while cutting the reference scan time by more than half (37\,s instead of 86\,s). Repeating the reference acquisition itself gives an MAE of about 0.1, a practical lower limit in vivo. Lower absolute MAE was achievable with longer multi-shot designs closer to $t_{\mathrm{ref}}$, illustrating the intended accuracy--duration trade-off when the target itself is a multi-shot image of the same subject, acquired on the same scanner in the same session with an established sequence, rather than an ideal signal equation.

\section{Discussion}
In this work we gained four main insights:
\begin{enumerate}
\item Coding skills of LLMs are sufficient to create valid and running PyPulseq sequence code, but sequences remain imperfect and performance fluctuates even with more detailed PyPulseq context.
\item An agent-based approach (Agent4MR) with structured validation feedback consistently yields correct, physically valid MR sequences with a single \investigator\ interaction across the three tested models, indicating that validation tools are more important than the underlying LLM.
\item Autonomous MR research (MR autoresearch) is in principle possible with such specialized agents given a well defined leaderboard challenge.
\item The same agent loop works in vivo: agents can send Pulseq sequences, reconstruct multi-coil data, and improve against a measured reference.
\end{enumerate}

While such agents are not immune to hallucinations, and often still follow non-ideal paths, the possibility of validation, using our suggested tools, strongly reduces the risk of the LLM returning a suboptimal sequence as solution.

\subsection*{Agent4MR harness}

The software around the model, the context, the checks, and the loop that runs them is called a \emph{harness}: in the case of Agent4MR it is the coding context and MR validation tools, as well as the use of a full MRI simulation, and the orchestration of these tools. How this harness is designed - e.g. avoiding information overload and enabling on-demand use of feedback - strongly influences how reliably the agent corrects physical and implementation errors. An early version of Agent4MR had too long text for the feedback, exceeding LLM context windows and degrading performance. 

For MRI sequence programming specifically, prior work is limited to our own GPT4MR and LLM4MR studies\cite{zaiss2024gpt4mr,zaiss2025llm4mr}, which established that general-purpose LLMs can act as coding assistants for PyPulseq when given structured prompts and error feedback. SeqGPT\cite{hussain2025seqgpt} represents a different paradigm: a model trained on pseudo-random MRI sequence data in the gammaSTAR language to generate pulse sequence structure for subsequent, conventional optimization by users or LLMs. An alternative line of work uses a domain-specific language for scanner-independent sequences and applies AI-driven optimization on top\cite{hoinkiss2023dsl}. More focused on education, but still combining MR physics and Pulseq knowledge, is the open-source tool PulsePal\cite{moskwa2025pulsepal}. To our knowledge, Agent4MR is the first agentic system that combines an LLM with automated execution, physics-based validation, and simulation in a single loop to produce scanner-ready sequences with minimal \investigator\ intervention. Beyond MRI sequence programming, a similar autoresearch paradigm has recently been demonstrated for CT reconstruction, where an LLM agent autonomously implemented and optimized reconstruction methods in a closed loop of coding, execution, and evaluation\cite{maier2026ctautoresearch}.

\subsection*{Limitations, Observations and practical notes}

In practice, we observed that Gemini replayed answers: identical prompts yielded identical outputs across separate sessions, whereas a slight rewording of the prompt gave a different answer. Whether this is caused by output caching or by near-deterministic sampling, we cannot tell; in either case, it is not an update of the model weights. The five repetitions were separate sessions, and the prompt wording was changed slightly when an identical answer reappeared. 

While the results above were limited to the spin-echo EPI sequence, we tested different standard sequence classes as well. Besides spin-echo EPI, the agent also produced a gradient-echo sequence, and it changed an existing gradient-echo implementation from one echo to several, with bipolar or monopolar readout gradients. In the autoresearch runs, agents also tried spiral readouts and a turbo spin echo (Figure~\ref{fig:MR_autoresearch_g31}), and in other runs GRASE implementations. 

The limitation of this study for the human benchmark is obvious, as we only recruited one student assistant for this. However, the condition to have a person that does not know the sequence, but at the same time has some experience in PyPulseq coding narrows down the number of potential candidates. Moreover, a 2.5\,h coding session is also a quite extensive participation when compared to typical survey times for scientific studies. Repeated runs, as done for the models, were not possible for the human, because a person cannot be reset to a context-free state; repeated sessions would measure learning rather than run-to-run variability. To benchmark humans and agents on a level playing field, a dedicated challenge such as a ``Pulseq Olympiad'' would be highly valuable.

\subsection*{MR autoresearch}
Karpathy's autoresearch approach\cite{karpathy2026autoresearch} forms a very interesting optimizer. While conventional optimization is common in MR sequence development, it is often limited to parts of the sequence design that can be formulated in a differentiable manner, meaning an analytical equation exists, and certain variables of the sequence are optimized. Even with more flexible optimizers that use discrete differences or gradient-free approaches, or reinforcement learning approaches, one needs to define the parameters that need to be optimized and at least the sequence skeleton containing these parameters. Now agent-based autoresearch has all flaws of hallucinating LLMs, but is completely free of any restriction of the to be optimized sequence. In each "iteration" the agent can create a new sequence skeleton and run specific optimizers for exactly this sequence situation. Thus, this provides a kind of framework for optimization and discovery of suitable sequences.

Very crucial is the definition of a loss or quality metric to provide a clear goal for the challenge. Pixel-wise MAE misses localized ghosting and distortion, so a low score is not the same as a preferred clinical image. We used MAE because mean squared error rewarded blur, for example by heavy filtering or by lowering the resolution. We stayed with MAE because we wanted the agents to resolve distortions explicitly; perceptual losses would have tolerated distortion to some extent. The principal idea of MR autoresearch shown here is independent of the loss metric, and different, more complex losses can be defined, matching the task or target to be achieved. The re-ranking with other metrics (Figure~\ref{sfig:3}) shows that the metric matters most for imperfect solutions, where it decides which compromise between distortion, blurring and ghosting is preferred. The solutions that came close to the target, however, were best in all tested metrics. Thus, the metric mainly alters the optimization trajectory, but this can be critical: an inappropriate metric may discard interesting solutions early. However, the investigator or another agent (with different metrics) can always intervene, e.g. ``explore in vivo entry 003 (Figure~\ref{fig:realscan_autoresearch}) further''. While this is very similar to classical optimization, it remains an important step and is a big degree of freedom for the \investigator. Comparing this with human researchers, goals or quality metrics often change during a detailed investigation, or further loss or regularization terms are added when necessary. Having the lowest loss might not be the most diagnostic, or most robust, or most general sequence, it simply is best with regard to this specific loss. Of course combined losses can be formed, yet their respective weighting then remains open.

With an agent that reliably generates and validates any possible pulse sequence, it becomes feasible to produce and simulate many different sequence variants in order to address a concrete clinical or methodological goal - for example, reducing a specific artifact or satisfying time, energy, or performance constraints. With MR autoresearch using Agent4MR we started to explore such trade-offs; our results point to a path where the agent supports systematic sequence optimization for well-posed objectives.

Closing the loop on the scanner further strengthens this view. The in vivo experiments on a whole-body scanner show that agents can operate with measured multi-coil data and a measured reference, not only with numerical phantoms.

\section{Conclusion}

We have shown that an agent-based framework (Agent4MR) with structured validation feedback and MR- and PyPulseq-specific context consistently produces correct, physically valid spin-echo EPI sequences with a single \investigator\ interaction across multiple base LLMs, requiring fewer interactions than an unprompted baseline and, in a single comparison, than a recently trained human with the same tools and task.  Beyond that benchmark, MR autoresearch on a FLAIR spin-echo EPI task limited to 10\,s showed that competing agents can refine both the pulse sequence and the reconstruction against a signal-equation target. The same approach also worked in vivo against a measured multi-shot reference. Agentic sequence development is therefore a promising path for reducing the effort and expertise required to implement and iterate on MRI pulse sequences, including open-ended, objective-driven exploration of acquisition and reconstruction.

\section*{Data Availability Statement}
The code, agent instructions and skills, as well as all experiments, run logs, leaderboard and figures of the MR autoresearch challenge shown in Figure~\ref{fig:MR_autoresearch_g31} and the simulation part of the in vivo challenge (Figure~\ref{sfig:4}) are openly available at \url{https://github.com/MRsources/mr_autoresearch}.

\section*{Acknowledgments}
This work was supported by the German Federal Ministry of Research, Technology and Space (BMFTR), grant number 16SV9585.

\bibliographystyle{unsrt}
\bibliography{literature}

\onecolumn
\section*{Supporting Information}

\renewcommand{\thefigure}{S\arabic{figure}}
\renewcommand{\theHfigure}{S\arabic{figure}}
\setcounter{figure}{0}

\subsection*{MR autoresearch}
\begin{figure}[htbp]
  \centering
  \includegraphics[width=1\textwidth]{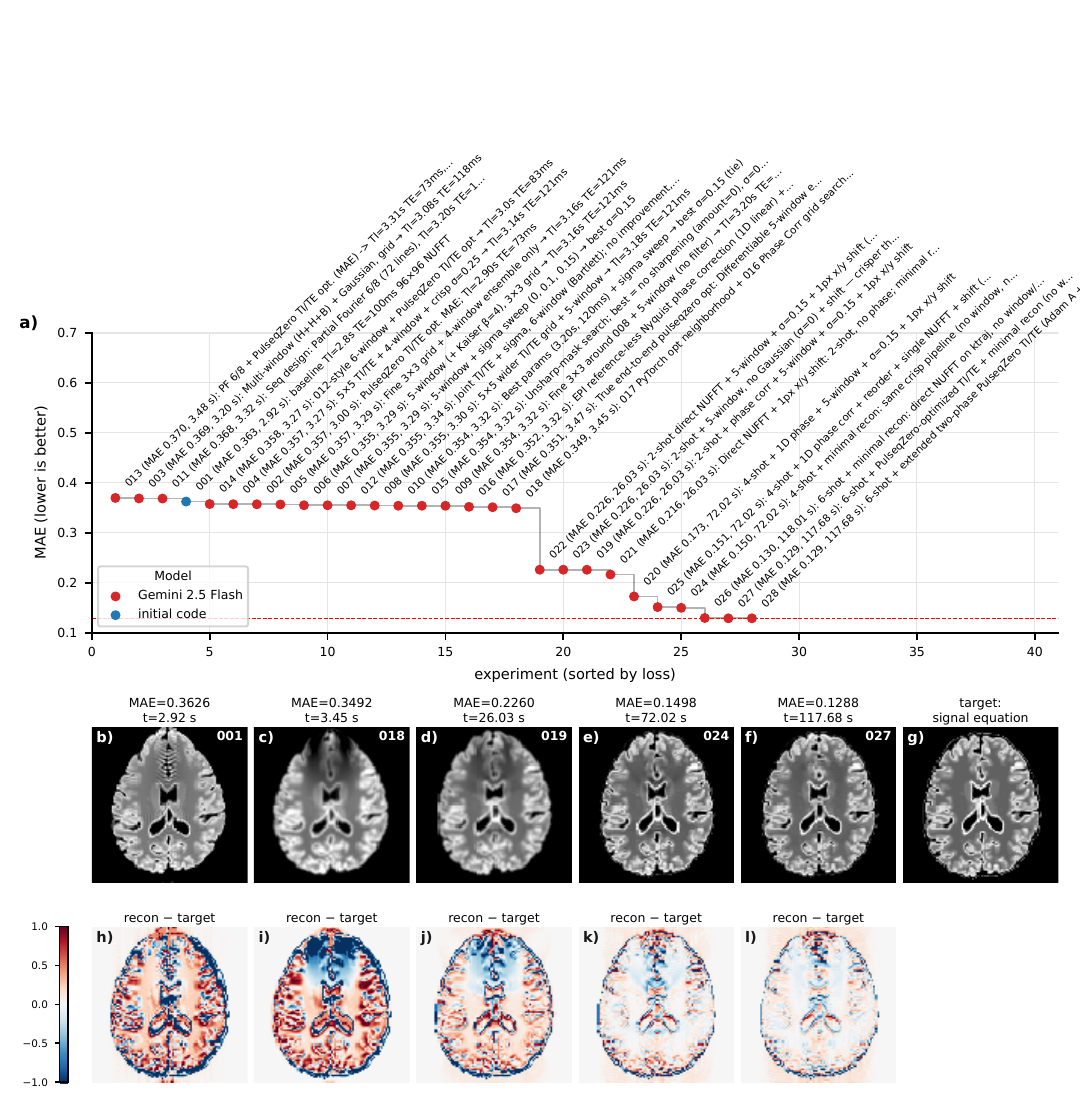}
  \caption{IR-SE-EPI MR autoresearch with the same setup as Figure~5 of the main text, but without time restriction, yet a strong $B_0$ inhomogeneity, leading to higher base MAE of 0.3626. After a long period of tuning the reconstruction, the agents come up with 2-shot and then 4-shot and 6-shot EPIs with much better performance. Panels as in Figure~5: (a) leaderboard, (b--f) reconstructions of the baseline, the best single-shot, 2-shot, 4-shot and 6-shot entries, (g) target, (h--l) difference images.}
  \label{sfig:1}
\end{figure}

In addition to the result presented in Figure~5 of the main text, we did a similar initial task of an IR SE EPI without time restriction, and stronger $B_0$ inhomogeneity. In Figure~\ref{sfig:1} we see that this leads to development of a very good match with the signal equation target by using a 6-shot EPI with additional MR-zero parameter optimization. However, this sequence is much longer than the initial single shot sequence, which is why we changed the task to an IR-SE-EPI with a scan time of at most 10\,s (Figure~5 of the main text).

\begin{figure}[htbp]
  \centering
  \includegraphics[width=1.0\textwidth]{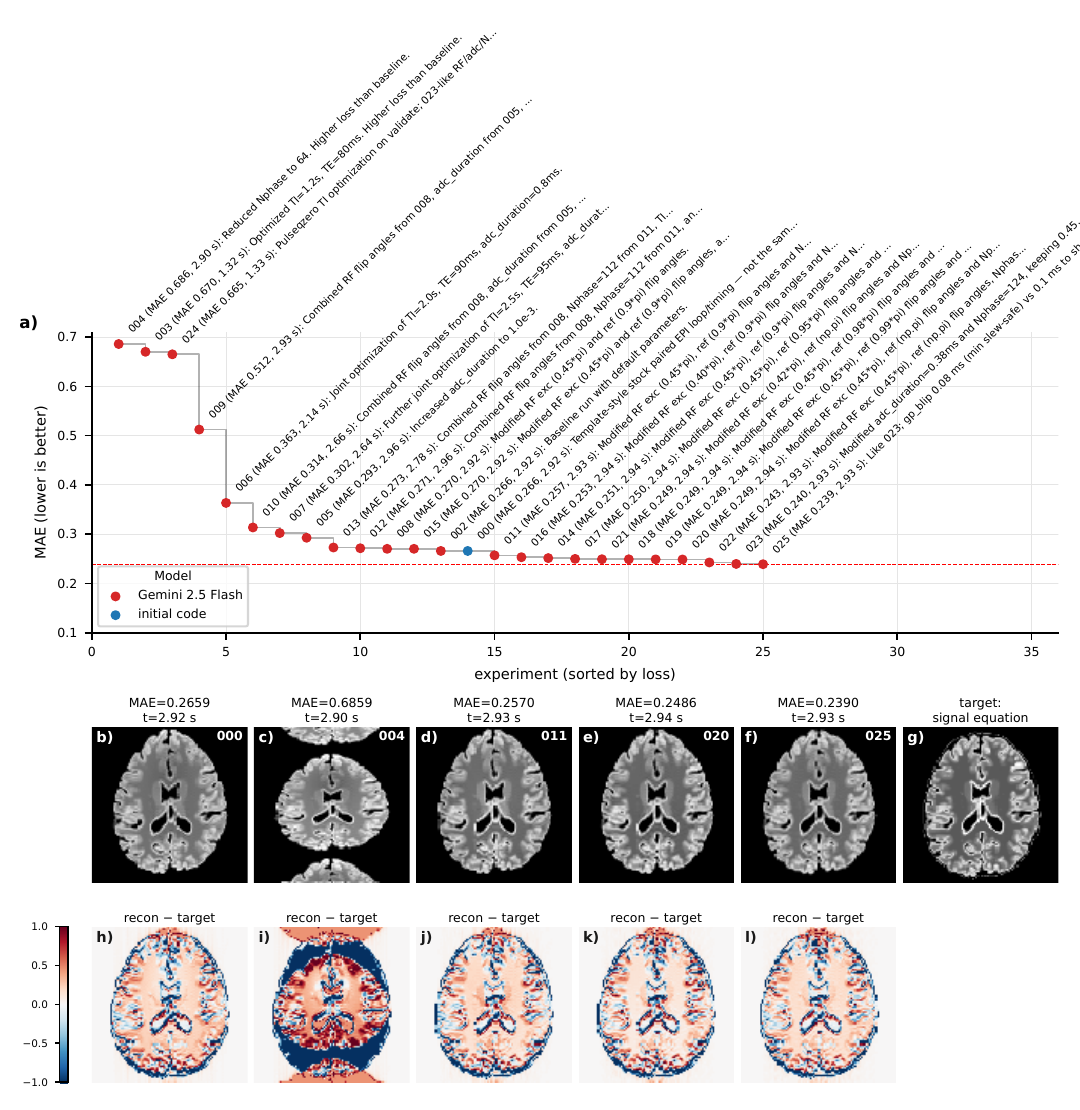}
  \caption{IR-SE-EPI MR autoresearch with the same setup as Figure~5 of the main text, but for the earlier Gemini model 2.5 Flash. Panels as in Figure~5: (a) leaderboard, (b--f) reconstructions, (g) target, (h--l) difference images.
  The overall performance improves with the model generation, indicated by the minimal MAE of each autoresearch experiment dropping from 0.239 (Gemini 2.5 Flash, this figure) to 0.167 (Gemini 3 Flash, Figure~5 of the main text) and to 0.135 for the spin-echo EPI of Gemini 3.1 Pro (Figure~6 of the main text).}
  \label{sfig:2}
\end{figure}

We also tested different model generations for the IR-SE-EPI with a scan time of at most 10\,s; as shown in Figure~\ref{sfig:2} for the earlier Gemini 2.5 Flash and in Figures~5 and 6 of the main text for Gemini 3 Flash and Gemini 3.1 Pro, the performance of the agents improves with the base model generation.

\textbf{Gemini 2.5 Flash (Figure~\ref{sfig:2})}

Starting from a $\sim$0.27 MAE baseline (000), early runs often hurt the score—aggressive TI/TE or MR-zero tuning (003–007, 024), coarser phase encoding (004, 009), or validate-only optimization—usually by tuning the wrong lever (sequence vs recon). Gains came from systematic sweeps of excitation/refocusing angles, Nphase, and adc\_duration (008–022). The best result is 025, using 81$^{\circ}$excitation and 180$^{\circ}$ refocusing—reduced flip angles that partially compensate $B_1^+$ inhomogeneity, together with adc\_duration $\approx$ 0.38 ms, Nphase = 124, and a shorter min slew-safe blip (0.08 ms vs 0.1 ms) to tighten EPI echo spacing.

\textbf{Gemini 3 Flash (Figure~5 of the main text)}

From the $\sim$0.27 MAE baseline (000/001), the first clear gains are readout / partial-Fourier / echo-timing tweaks plus MR-zero on TI/TE/FA (002–004), into the $\sim$0.22–0.24 range. That is combined with two-shot segmented / partial-Fourier EPI and joint MR-zero over timings and shot scaling (005–008), near $\sim$0.21 MAE. The largest jump uses almost the full 10 s: long recovery, interleaved 2-shot coverage, MR-zero, and iterative recon (009–011), around $\sim$0.17–0.18. Iterative Recon refinements, TV, multi-group / linear phase correction—yield the best $\sim$0.1666 MAE (016/017).

\textbf{Gemini 3.1 Pro (Figure~6 of the main text)}

The first clear improvement is tighter EPI echo spacing (004, about 0.23). Most of the progress comes from long, systematic work on two-shot asymmetric and partial-Fourier EPI, including ramp-aware density and shot layout (012–020) using the full $\sim$10 s. Spiral readout (021) fails and is no longer considered; dual spin-echo (022) is mainly a short $\sim$3 s scan at $\sim$0.24 MAE. The real breakthrough is explicit $B_0$ correction: grid unwarping (023) then polynomial $B_0$ with shot-2 scaling and phase (024, about 0.159). Applying similar $B_0$ recon to a minimal single-shot (025) does not transfer (about 0.38).

In the continued run (026–033), the agents further refined the two-shot asymmetric 96$\times$64 EPI with 8$\times$8 grid unwarping and k-space phase correction, giving the best spin-echo EPI (028, MAE about 0.135, 9.92\,s). They then left the EPI concept and switched to a single-shot RARE (turbo spin echo) readout with centric-like phase encoding, an echo spacing of 4.0\,ms and TE\,=\,100\,ms (031, about 0.079, 3.19\,s). An inversion-recovery version with centric ordering (032, about 0.053, 3.10\,s) and a version with a shorter echo spacing of 3.08\,ms and an analytical contrast adjustment (033, about 0.044, 3.10\,s) gave the lowest MAE of all runs, at about a third of the allowed scan time.

\clearpage
\subsection*{Dependence on the image metric}
\begin{figure}[htbp]
  \centering
  \includegraphics[width=1.0\textwidth]{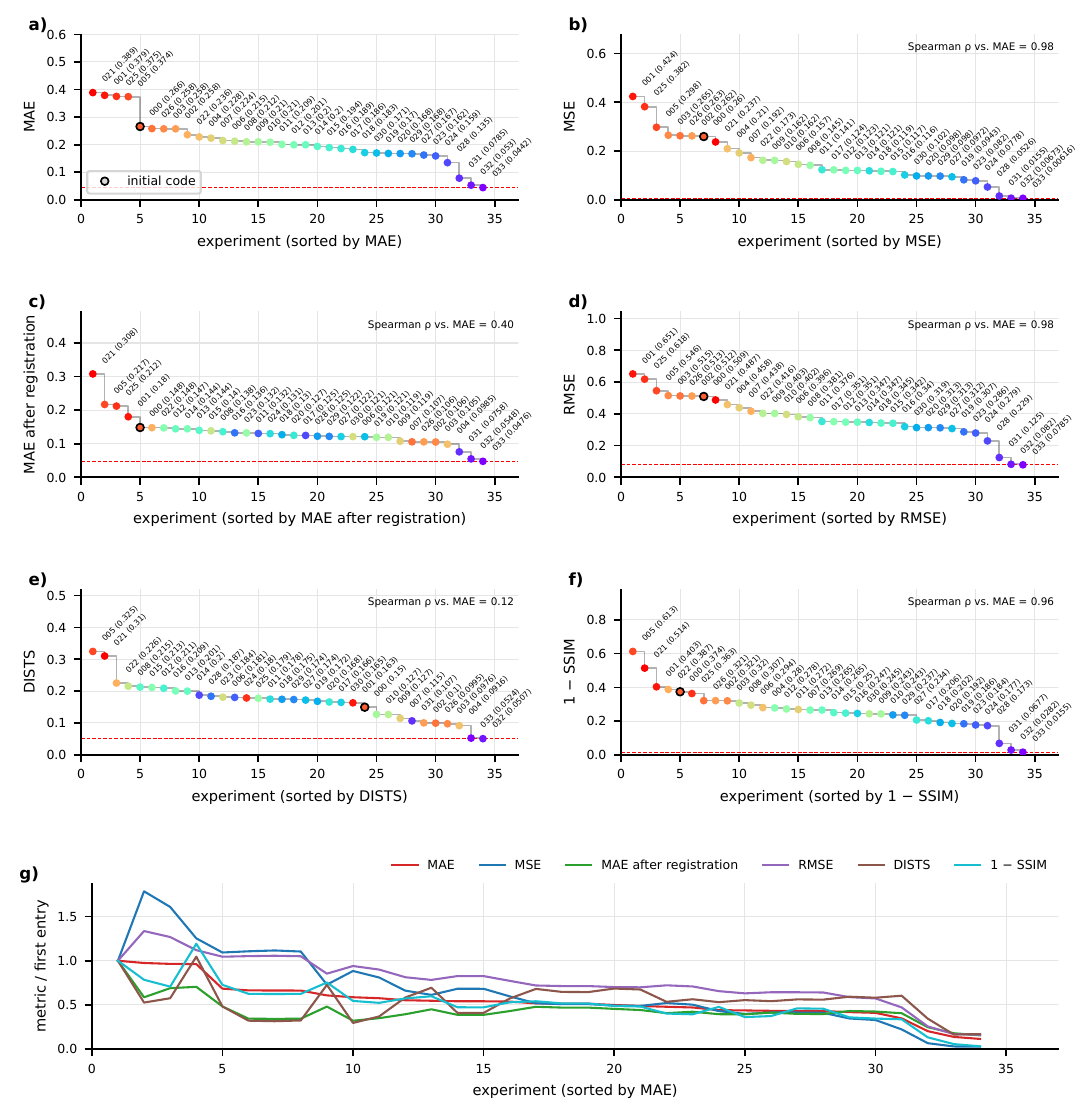}
  \caption{Gemini 3.1 Pro MR autoresearch (Figure~6 of the main text) ranked by different image metrics between reconstruction and signal-equation target: (a) MAE, the challenge loss (leaderboard values, as Figure~6a), (b) MSE, (c) MAE after deformable registration of each reconstruction to the target, (d) RMSE, (e) the perceptual metric DISTS and (f) $1-\mathrm{SSIM}$; each panel is sorted by its own metric and gives the Spearman rank correlation with the MAE ranking. In (a--f), each experiment keeps the colour of its MAE rank (rainbow from red, worst MAE, to violet, best MAE), so colour jumps reveal changes in the ranking; the initial code is marked by a black edge. (g) All metrics plotted over the MAE ranking of (a), each normalized to its value for the initial code (experiment 000).}
  \label{sfig:3}
\end{figure}

To test how much the results depend on the choice of MAE as challenge loss, all Gemini 3.1 Pro experiments (Figure~6 of the main text) were re-ranked by five further metrics between reconstruction and signal-equation target (Figure~\ref{sfig:3}). MSE and RMSE are voxel-wise like MAE, but weight large errors more strongly; $1-\mathrm{SSIM}$ measures structural similarity in local windows. To separate geometric distortion from contrast errors and artifacts, each reconstruction was deformably registered to the target (TV-L1 optical flow) and the MAE was computed after registration. Finally, the perceptual metric DISTS (Ding et al., IEEE TPAMI 2022) compares deep (VGG) image features and is designed to tolerate small geometric deviations. These metrics were recomputed from the stored target and reconstruction images of each experiment; the MAE recomputed this way agrees with the leaderboard values to within 0.003.

The metrics judge the intermediate EPI solutions differently. MSE, RMSE and $1-\mathrm{SSIM}$ follow the MAE ranking closely (Spearman rank correlation with the MAE ranking of 0.96--0.98). The MAE after registration deviates strongly (0.40): most EPI entries reach a similar value of about 0.1--0.15 once distortions are removed, i.e.\ a large part of the MAE gain of the EPI variants is a reduction of geometric distortion. The DISTS ranking deviates most (0.12) and rates many tuned EPI variants worse than the initial code (Figure~\ref{sfig:3}g). These intermediate entries are exactly the compromises between distortion, blurring and ghosting discussed for Figure~5c and 5e of the main text, and which compromise is preferred depends on the metric. Note that the deformable registration can also partly compensate blurring, so the MAE after registration is a lower bound of the non-geometric error.

Once an image comes close to the target, however, all metrics converge. The single-shot TSE entries 032 and 033 are the two best entries in every metric: 033 is best in five of the six metrics, and 032 is best in DISTS with 033 a close second. Compared with the initial code, all metrics are reduced to 2--37\%. Thus, the metric decides the ranking among imperfect solutions, but a solution close to the target was best in all tested metrics.

\subsection*{In silico counterpart of the in vivo MR autoresearch}

\begin{figure}[htbp]
  \centering
  \includegraphics[width=1.0\textwidth]{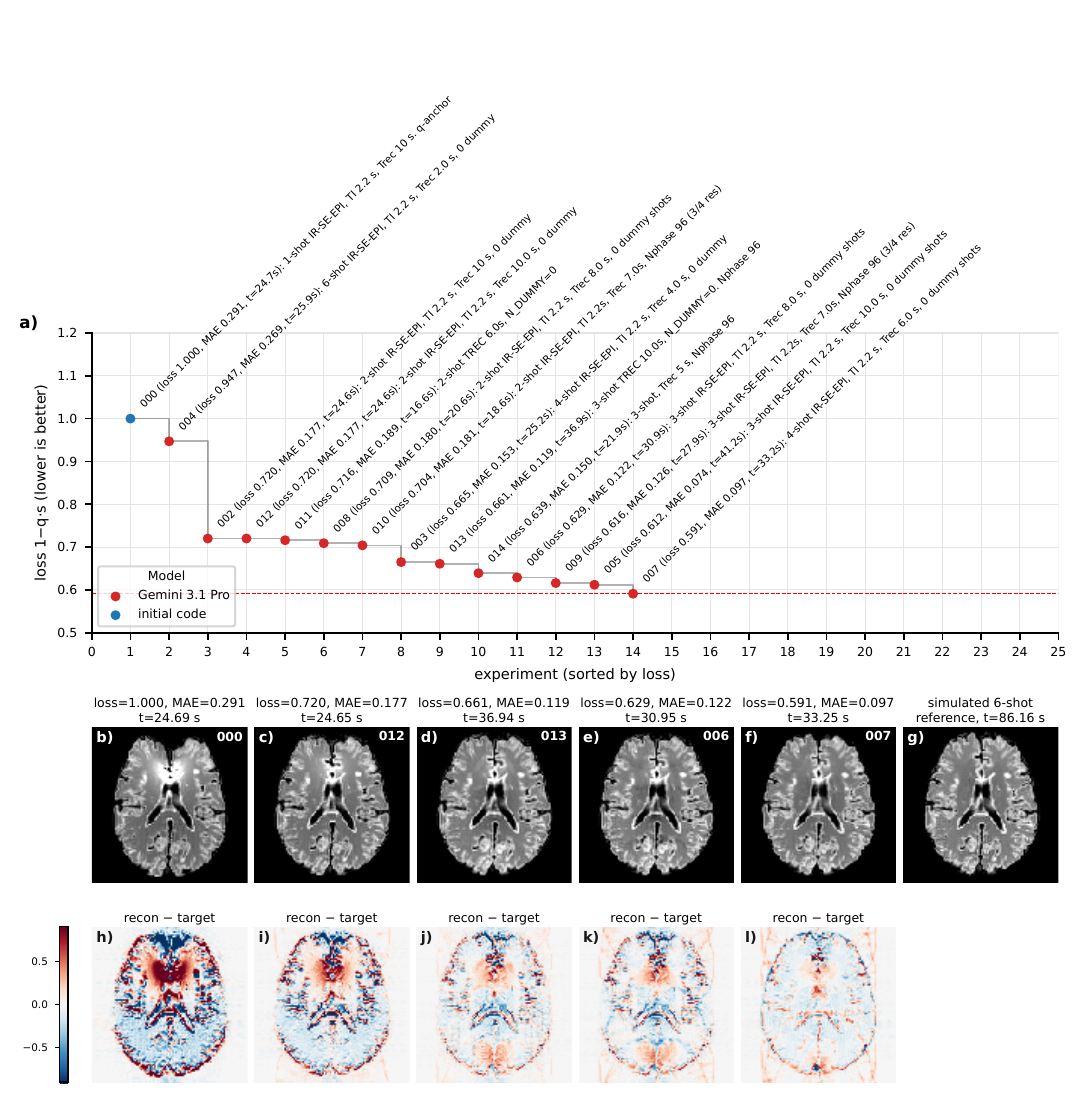}
  \caption{In silico counterpart of the in vivo MR autoresearch (Figure~8 of the main text): the 14 candidates of the in vivo session simulated with MR-zero (single receive channel, brain phantom without fat) and scored against the simulated 6-shot reference. The $\Delta B_0$ map of the phantom was scaled by a factor of 4, which creates a similarly poor shim as in the in vivo scan. Ranking uses $\mathrm{loss}=1-q\cdot s$ with $\mathrm{MAE}_{0}=0.291$ (simulated baseline 000) and $t_{\mathrm{ref}}=86.16$\,s (lower is better).
  (a)~Leaderboard progress: experiments sorted by decreasing loss.
  (b--f)~Reconstructions of the same experiments as in Figure~8, sorted by decreasing loss in simulation: baseline single-shot IR-SE-EPI 000, 2-shot 012, 3-shot 013 (best in vivo), 3-shot 006 and 4-shot 007 (best in simulation).
  (g)~Simulated 6-shot reference, $t_{\mathrm{ref}}=86.16$\,s.
  (h--l)~Difference images (recon $-$ target) of (b--f).}
  \label{sfig:4}
\end{figure}

For Figure~\ref{sfig:4}, all 14 candidates of the in vivo session (Figure~8 of the main text; experiments 000 and 002--014) and the 6-shot reference (001) were simulated with MR-zero on the numerical brain phantom (subject~04 at 3\,T without fat, slice at $z=0$\,mm) with a single receive channel. The measured $\Delta B_0$ map of the phantom was scaled by a factor of 4 (maximum $|\Delta B_0|$ in the slice 173\,Hz instead of 43\,Hz) to create a similarly poor shim as in the in vivo scan. The sequences and reconstructions of the candidates were used unchanged, i.e.\ the same NUFFT reconstruction including the navigator-based odd/even phase correction as in vivo. Every candidate was scored against the simulated 6-shot reference with the loss $1-q\cdot s$ of the main text, using $t_{\mathrm{ref}}=86.16$\,s and the simulated MAE of the single-shot baseline 000 ($\mathrm{MAE}_{0}=0.291$).

In simulation, the 4-shot IR-SE-EPI 007 ($T_{\mathrm{rec}}=6$\,s) was the best entry (MAE\,=\,0.097, $t=33.25$\,s, loss\,=\,0.591), followed by the 4-shot 005 (loss\,=\,0.612) and the 3-shot entries 009 and 006 (0.616 and 0.629). The in vivo winner 013 ranked sixth (MAE\,=\,0.119, loss\,=\,0.661). With the strong off-resonance, the single-shot baseline 000 suffers the strongest distortion and is the worst entry, and all multi-shot candidates improve on it. The simulation contains no subject motion, noise or other system imperfections, and thus isolates the effect of the sequence parameters on distortion and contrast, whereas in vivo these additional error sources determine a large part of the MAE and the ranking; in particular, 007 was the worst entry in vivo.

\renewcommand{\tablename}{Supporting Table}
\renewcommand{\thetable}{S\arabic{table}}
\renewcommand{\theHtable}{S\arabic{table}}
\setcounter{table}{0}

\begin{table}[htbp]
  \centering
  \caption{Prompts used in the experiments of the main text.}
  \label{stab:prompts}
  \begin{tabular}{p{0.25\textwidth}p{0.68\textwidth}}
    \hline
    Experiment & Prompt \\
    \hline
    Spin-echo EPI, LLM and LLM4MR setups & ``Please, read the instructions. Code a spin echo EPI sequence (64$\times$64, FOV\,=\,(0.2, 0.2, 0.008)\,m, TE\,=\,100\,ms).'' \\[0.4em]
    Spin-echo EPI, Agent4MR setup & ``Please, read the Agent4MR instructions. Code a spin echo EPI sequence (64$\times$64, FOV\,=\,(0.2, 0.2, 0.008)\,m, TE\,=\,100\,ms). Finish only when you are 100\% sure it is properly implemented. Use the terminal to test the codes. Use all tools to validate the sequence.'' \\[0.4em]
    MR autoresearch (simulation and in vivo) & ``Please, read the instructions. Approach to win the challenge!'' The challenge instructions were provided as a Markdown file. \\
    \hline
  \end{tabular}
\end{table}

\end{document}